\documentclass[letter,twocolumn]{jpsj3}
\usepackage{txfonts}
\usepackage{bm}
\usepackage{mathrsfs}
\usepackage{here}
\usepackage{color}
\usepackage{graphicx}
\newcommand{\simg}{\stackrel{>}{_\sim}}
\newcommand{\siml}{\stackrel{<}{_\sim}}

\title{
Strong Correlation Effect and In-gap State in the Doped Electron-Hole Two-Band Hubbard Model Based on the Dynamical Mean-Field Theory
}

\author{Masashi Akiyama$^{1}$\thanks{m.akiyama.phys@gmail.com}, Yusuke Inokuma$^{2}$, and Yoshiaki \textsc{\=O}no$^{1}$}
\inst{$^1$Department of Physics, Niigata University, Niigata, 950-2181, Japan\\
$^2$Nuclear Science and Engineering Center, Japan Atomic Energy Agency, Ibaraki, 319-1195, Japan\\} 

\abst{
We investigate the strong correlation effect in the spinless electron-hole two-band Hubbard model using the dynamical mean-field theory. At half filling, both the renormalization factor $Z$ and the number of conduction electrons (valence holes) $n_c$ decrease with increasing the interband Coulomb interaction $U$ down to $Z\sim 0.15$ and $n_c\sim 0.1$ for $U_c \sim \mbox{bandwidth}$ at which the first-order Lifshitz transition occurs from a correlated semimetal with a large effective mass $m^*/m=Z^{-1}$ to a band insulator with a finite gap due to the Hartree shift. A slight hole doping $x$ in the band insulator with a large $U>U_c$ yields a remarkable correlated semimetal with $Z\sim 0.1$ at $x\sim 0.01$, where in-gap states emerge above the valence band top and those weights increase with increasing $x$ together with the increase in $Z$ similar to the in-gap states observed in doped Mott insulators. 
}

\begin{document}
\maketitle

Excitonic insulator (EI), which is an ordered state where electron-hole pairs (excitons) undergo Bose-Einstein condensation (BEC) at low temperatures due to interband Coulomb interaction in the semimetal-semiconductor transition regime, was originally proposed in the 1960s\cite{Keldysh,Cloizeaux,Jerome,Halperin} and has recently been extensively investigated especially for several new candidate materials\cite{Kaneko3}. 
Among them, Ta$_2$NiSe$_5$ shows an orthorhombic-to-monoclinic phase transition at $T_c$=328 K, below which the flattening of the valence band top is observed in the ARPES experiments\cite{Wakisaka} and is well accounted for by the excitonic condensation from a normal semiconductor to the EI on the basis of the BCS type mean-field theory\cite{Kaneko}. The strong correlation effects which are considered to be important in the BEC and BEC-BCS crossover regimes have also been investigated in the one-dimensional spinless two-band Hubbard model (or extended Falicov-Kimball model) by using the density-matrix renormalization group (DMRG)\cite{Sugimoto} and have been found to yield a novel insulator state which is responsible for the optical conductivity spectra in Ta$_2$NiSe$_5$\cite{Larkin}.

More recently, carrier doping effects on Ta$_2$NiSe$_5$ have been intensively studied by using the elemental substitutions\cite{Hirose,Tsuchida}, and metallic behaviors have been successfully observed in (Ta$_{1-x}$Ti$_x$)$_2$NiSe$_5$ with $x>0.06$ where holes are considered to be doped in the EI\cite{Tsuchida}. In the metallized Ta$_2$NiSe$_5$, the Sommerfeld coefficient has been observed to show a large effective mass\cite{Tsuchida2} indicating importance of the strong correlation effects in the hole doped EI. Theoretically, doping effects in the one-dimensional spinless two-band Hubbard model have recently been investigated by using the DMRG\cite{Kaneko2} and the emergence of a correlation-driven in-gap branch due to the doping has been observed, similar to that of doped Mott insulators. The correlation effects such as the mass enhancement, however, were not discussed there as the one-dimensional systems are described by the Tomonaga-Luttinger liquid instead of the Fermi liquid  as observed in the metallized Ta$_2$NiSe$_5$.

In this Letter, we investigate the spinless electron-hole two-band Hubbard model (or extended Falicov-Kimball model) on the basis of the dynamical mean-field theory (DMFT)\cite{Georges}, which becomes exact in infinite dimensions and is expected to be a good approximation in three dimensions, so as to take into account the strong correlation effects including the excitonic correlations. The DMFT has already been adopted for studying the EI in the similar models\cite{Kunes1,Kunes2,Sasaki}. However, the strong correlation effects on the metallic state especially in the doped case were not discussed there. The purpose of this letter is to elucidate the strong correlation effect on the normal state without excitonic condensation especially focus on the hole doped case, as discussed in the doped Mott insulator without antiferromagnetic order\cite{Georges}.

We consider the spinless electron-hole two-band Hubbard model given by the following Hamiltonian\cite{Sugimoto,Kaneko2,Sasaki}: 
\begin{equation}
H=\sum_{i,j}t_{ij}(c_{i}^\dagger c_{j} - f_{i}^\dagger f_{j}) 
+\Delta\sum_{i}n_{i}^{c}+U\sum_{i}n_{i}^{c}n_{i}^{f},
\end{equation}
where $c_{i}^\dagger$ ($c_{i}$) and $f_{i}^\dagger$ ($f_{i}$) are the creation (annihilation) operators for conduction ($c$) and valence ($f$) electrons at site $i$ and  $n_{i}^{c} = c_{i}^\dagger c_{i}$ and $n_{i}^{f} = f_{i}^\dagger f_{i}$, respectively. $t_{ij}$ ($-t_{ij}$) is the transfer integral between the sites $i$ and $j$ for $c$ ($f$) electrons, $\Delta$ is the atomic $c$ level with setting the atomic $f$ level to be zero resulting in the $c$-$f$ level splitting $\Delta$ and $U$ is the on-site $c$-$f$ (interband) Coulomb interaction. We assume that each site $i$ is on the Bethe lattice with the connectivity $z$ and $t_{ij}=t/\sqrt{z}$ between the nearest-neighbor sites $i$ and $j$, where the bare density of states (DOS) becomes semielliptic with the bandwidth $W=4t$ in the limit $z\to \infty$ as discussed in the DMFT.

To solve the model eq. (1), we use the DMFT \cite{Georges} where the self-energy $\Sigma(i\omega_{n})$ becomes local in infinite spatial dimensions (infinite connectivity $z$) and then the local Green's functions for $c$ and $f$-electrons $G_c(i\omega_{n})$ and $G_f(i\omega_{n})$ satisfy the following self-consistency conditions
\begin{flalign}
&{\cal{G}}_c(i\omega_{n})^{-1}=i\omega_{n}+\mu-\Delta-t^{2}G_c(i\omega_{n}), \\
&{\cal{G}}_f(i\omega_{n})^{-1}=i\omega_{n}+\mu-t^{2}G_f(i\omega_{n}),
\end{flalign}
where $\mu$ is the chemical potential and $\omega_n =(2n+1)\pi T$ is the fermionic Matsubara frequency with the temperature $T$. In eqs. (2) and (3), ${\cal{G}}_c(i\omega_{n})$ and ${\cal{G}}_f(i\omega_{n})$ are the Weiss functions which include effects of the interaction at all the sites except the local site and can be written by the noninteracting Green's functions $G_c^{0\rm{And}}(i\omega_n)$ and $G_f^{0\rm{And}}(i\omega_n)$ of the impurity Anderson model embedded in an effective medium given by
\begin{flalign}
&H_{And}=\epsilon^cn^c+\sum_{l=2}^{N_{s}}\epsilon^c_{l} a_{l}^{c \dagger}a_{l}^c 
+\sum_{l=2}^{N_{s}}V^c_{l}(c^{\dagger}a^c_{l}+a^{c \dagger}_{l}c) +Un^c n^f \nonumber \\
&\ \ \ \ \ \ \ \ +\epsilon^f n^f+\sum_{l=2}^{N_{s}}\epsilon^f_{l} a^{f \dagger}_{l}a^f_{l} 
+\sum_{l=2}^{N_{s}}V^f_{l}(f^{\dagger}a^f_{l}+a^{f \dagger}_{l}f),
\end{flalign}
where $a_{l}^\dagger$ ($a_{l}$) is the creation (annihilation) operator for electrons of the effective medium at site $l$, $\epsilon^{c(f)}$ is the impurity level and  $\epsilon^{c(f)}_l$ ($l=2,3,...,N_s$) are levels of the effective medium hybridized with the impurity by $V^{c(f)}_l$. Then, the impurity Green's functions $G_{c(f)}^{\rm{And}}(i\omega_n)$ in eq. (4) correspond to the local Green's functions $G_{c(f)}(i\omega_{n})$ in eqs. (2) and (3), and the parameters $\epsilon^{c(f)}$, $\epsilon^{c(f)}_l$, $V^{c(f)}_l$ ($l=2,3,...,N_s$) in eq. (4) are determined so as to make ${\cal{G}}_{c(f)}(i\omega_{n}) \ (G_{c(f)}^{0\rm{And}}(i\omega_n))$ and $G_{c(f)}(i\omega_{n}) \ (G_{c(f)}^{\rm{And}}(i\omega_n))$  satisfy the self-consistency conditions eqs. (2) and (3).

In the present study, we solve the effective impurity Anderson model eq. (4) with finite system sizes $N_{s}=4,$ 6 and 8 by using the exact diagonalization (ED) method\cite{Caffarel,Ono,Inokuma} and obtain $G_{c(f)}^{\rm{And}}(i\omega_n)$ from the eigenvalues and eigenvectors calculated with the use of the Householder method. Although the system sizes are quite small, the size dependence of the obtained numerical results is found to be quite small as previously observed in the single-band and the two-band Hubbard models\cite{Caffarel,Ono,Inokuma} and as explicitly shown in Figs. 1, 3 and 5 in the present model.  We set $t=1$ corresponding to order of 1eV in real materials, $T=0.02$ and $\Delta=0.5$ and we vary the parameters $U$ and the hole doping concentration which is given by $x\equiv 1-n_c-n_f$ with the occupation number of $c$ ($f$) electron $n_c$ ($n_f$) and is determined by choosing a suitable value of $\mu$ in eqs. (2) and (3). Due to the particle-hole symmetry of the model eq. (1), the results for the electron doping are essentially the same as those for the hole doping, and then we only show the latter results in this paper.

Taking into account the Hatree shift, the $c$-$f$ level splitting is enlarged as $\Delta+U(n_f-n_c)$ resulting in the band gap 
\begin{equation}
E_{\rm gap}=\Delta+U(n_f-n_c)-W. 
\end{equation}
Therefore, when $U$ increases at half-filling, a Lifshitz transition from a semimetal ($E_{\rm gap}<0$) to a band insulator ($E_{\rm gap}>0$) is expected to occur at $U_c \sim W-\Delta$. In the DMFT, in addition to the Hatree shift, the correlation effect around the transition is considered to be important especially in the case with a small $\Delta$ resulting in a large $U_c$. To explicitly discuss the correlation effect, we calculate the renormalization factor 
$Z_{c(f)}=\bigl[1-\frac{\rm{d} \rm{Re}\Sigma_{c(f)}(\omega)}{\rm{d} \omega}|_{_{\omega=0}}\bigr]^{-1}$ 
with $\Sigma_{c(f)}(\omega)={\cal{G}}_{c(f)}(\omega)^{-1}-G_{c(f)}(\omega)^{-1}$
for $c$ ($f$) electron. When the system is metallic, the mass enhancement of the $c$ ($f$) electron is given by $(m^*/m)_{c(f)}=Z_{c(f)}^{-1}$ which is largely enhanced in the strong correlation regime where $Z$ becomes small. To discuss the electronic states in more detail, we also calculate the spectral function of $c$ ($f$) electron 
$\rho_{c(f)}(\omega)=-\frac{1}{\pi}{\rm Im}G_{c(f)}(\omega+i0_+)$ which is a collection of delta functions for a finite $N_s$, and then we merge the results from $N_s=6$ and 8 to increase the data points and plot those with Lorentzian broadening with a width of 0.1eV for easy to see.

First, we investigate the correlation effect at half-filling without doping ($x=0$). 
Figures~1 (a) and (b) show the $U$ dependence of $Z_c$ and $Z_f$ and that of $n_c$ and $n_f$ together with $E_{\rm gap}$ given in eq. (5), where the $N_s$ dependence of the obtained results for $N_s=4$, 6 and 8 is quite small as mentioned before. As shown in Fig. 1 (a), where $Z_c=Z_f$ due to the particle-hole symmetry of the model eq. (1), $Z$ monotonically decreases with increasing $U$ down to $Z=0.15$ for $U=U_c=4.3$ at which $Z$ shows a jump to 0.65, and then $Z$ monotonically increases towards unity for $U>U_c$. Correspondingly, $n_c=1-n_f$ monotonically decreases with increasing $U$ down to $n_c=0.10$ for $U=U_c$ at which $n_c$ shows a jump to 0.05, and then $n_c$ further decreases towards zero for $U>U_c$. The results of $Z$ and $n_c$ together with the result of $E_{\rm gap}$ (see Fig.~1 (b)) indicate that the first-order Lifshitz transition takes place at $U=U_c$ from a correlated semimetal ($E_{\rm gap}<0$) with a large effective mass $m^*/m=Z^{-1}$ to a band insulator with a finite gap ($E_{\rm gap}>0$) due to the Hartree shift. We note that, in the band insulator for $U>U_c$, $Z$ is unrelated to $m^*/m$ but is considered to relevant to a correlation strength which decreases with increasing $U$ and becomes almost zero ($Z\sim 1$) for $U\simg 6$ where $n_c \sim 0$. 

\begin{figure}[t]
\begin{center}
\begin{center}
\includegraphics[width=6.4cm]{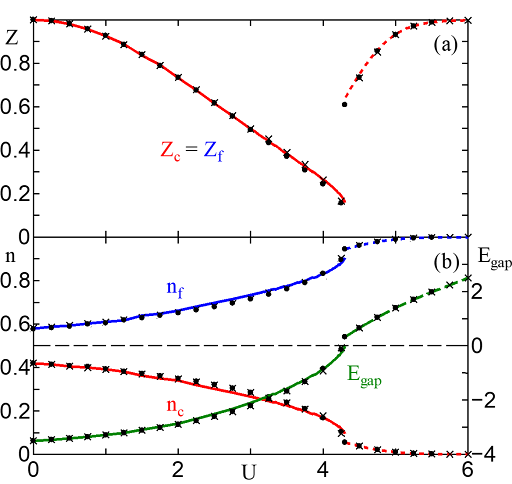}
\end{center}
\caption{\label{fig1} 
(Color online) The renormalization factor of $c$ ($f$) electron $Z_c$ ($Z_f$) (a) and the occupation number of $c$ ($f$) electron $n_c$ ($n_f$) together with the band gap $E_{\rm gap}$ as functions of the $c$-$f$ Coulomb interaction $U$  at half filling $x=0$ for $N_s =4$ (solid (dashed) lines for metallic (insulating) state), $N_s =6$ (dots) and $N_s =8$ (crosses).}
\end{center}
\end{figure}

\begin{figure}[t]
\begin{center}
\begin{center}
\includegraphics[width=6.4cm]{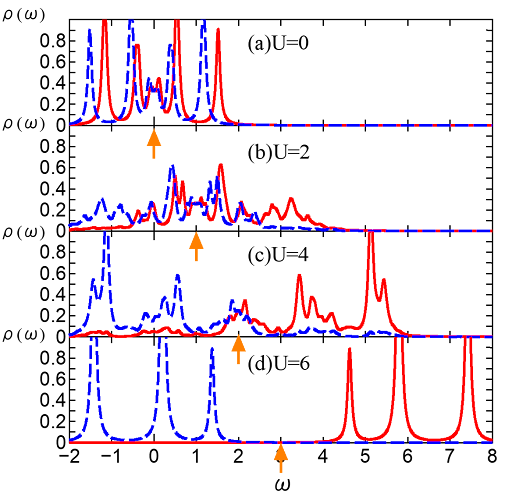}
\end{center}
\caption{\label{fig2} 
(Color online) The spectral function of $c$ electron $\rho_c(\omega)$ (solid lines) and those of $f$ electron  $\rho_f(\omega)$ (dashed lines) as functions of the frequency $\omega$ at half filling $x=0$ for $U=0$ (a),  $U=2$ (b), $U=4$ (c) and $U=6$ (d), where the arrows indicate the chemical potential $\mu$. The results from $N_s =6$ and 8 are merged and plotted with Lorentzian broadening with a width of 0.1eV.}
\end{center}
\end{figure}

Figs.~2 (a)-(d) show the $\omega$ dependence of the spectral function of $c$ ($f$) electron $\rho_{c(f)}(\omega)$ at half filling $x=0$ for several values of $U$, When $U$ increases, the $c$-$f$ level splitting increases and correspondingly the band overlapping in the semimetal given by $-E_{\rm gap}>0$ decreases for $U<U_c$ as shown in Figs.~2 (a)-(c), and then the system becomes a band insulator with a finite gap $E_{\rm gap}>0$ for $U>U_c$ as shown in Fig.~2 (d). We see that $\rho_{c(f)}(\omega)$ for $U=2$ and 4 in the correlated semimetal are largely deformed from the bare one for $U=0$, while $\rho_{c(f)}(\omega)$ for $U=6$ in the band insulator with $Z\sim 1$ is similar to the bare one except the enlarged band splitting. We note that the weight of the quasiparticle peak centered at $\mu$ in the semimetal decreases with increasing $U$ together with the decrease in $Z$, similar to the case with the Hubbard model at half-filling\cite{Georges} where the weight of the quasiparticle peak and $Z$ decrease with increasing $U$ towards zero at $U_c \sim \mbox{bandwidth}$ above which the system becomes the Mott insulator in contrast to the band insulator in the present model.

Next, we discuss the effect of the hole doping $x$ on the correlated semimetal for $U<U_c$. 
Figures~3 (a) and (b) show the $x$ dependence of $Z_c$ and $Z_f$ and that of $n_c$ and $n_f$ together with $E_{\rm gap}$ given in eq. (5) for $U=4$, where the $N_s$ dependence of the obtained results for $N_s=4$, 6 and 8 is quite small as seen in Fig.~1. When $x$ increases, the both $Z_c$ and $Z_f$ monotonically increase towards unity for $x=1$ where $n_c+n_f=1-x=0$. The result of $Z_c<Z_f$ for $x>0$ indicates that the correlation effect of $c$ electron is relatively larger than that of $f$ electron as $\mu$ moves away from (closer to) the atomic $c$ ($f$) level and then the charge transfer energy of $c$ electron is relatively larger than that of $f$ electron (see Fig.~4). As shown in Fig.~1 (b), $n_f$ ($n_c$) steeply (gradually) decreases with increasing $x$ towards zero for $x=1$, and then the Hartree shift decreases with increasing $x$ resulting in the increase in the band overlapping $-E_{\rm gap}$ up to $W-\Delta=3.5$ for $x=0$.

\begin{figure}[t]
\begin{center}
\includegraphics[width=6.4cm]{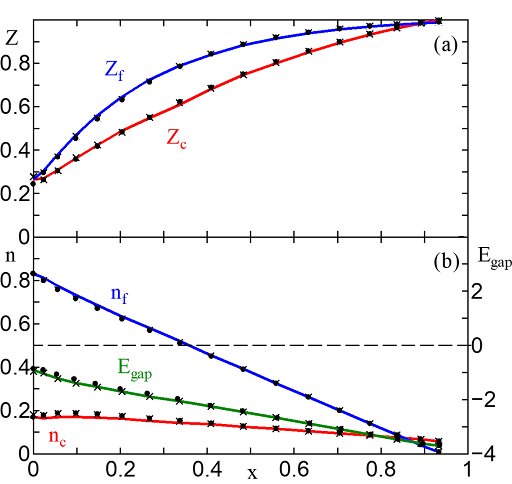}
\end{center}
\caption{\label{fig3} 
(Color online) The renormalization factor of $c$ ($f$) electron $Z_c$ ($Z_f$) (a) and the occupation number of $c$ ($f$) electron $n_c$ ($n_f$) together with the band gap $E_{\rm gap}$ as functions of the hole doping $x$ at $U=4$ for $N_s =4$ (solid line for metallic state), $N_s =6$ (dots) and $N_s =8$ (crosses).}
\end{figure}

\begin{figure}[t]
\begin{center}
\includegraphics[width=6.4cm]{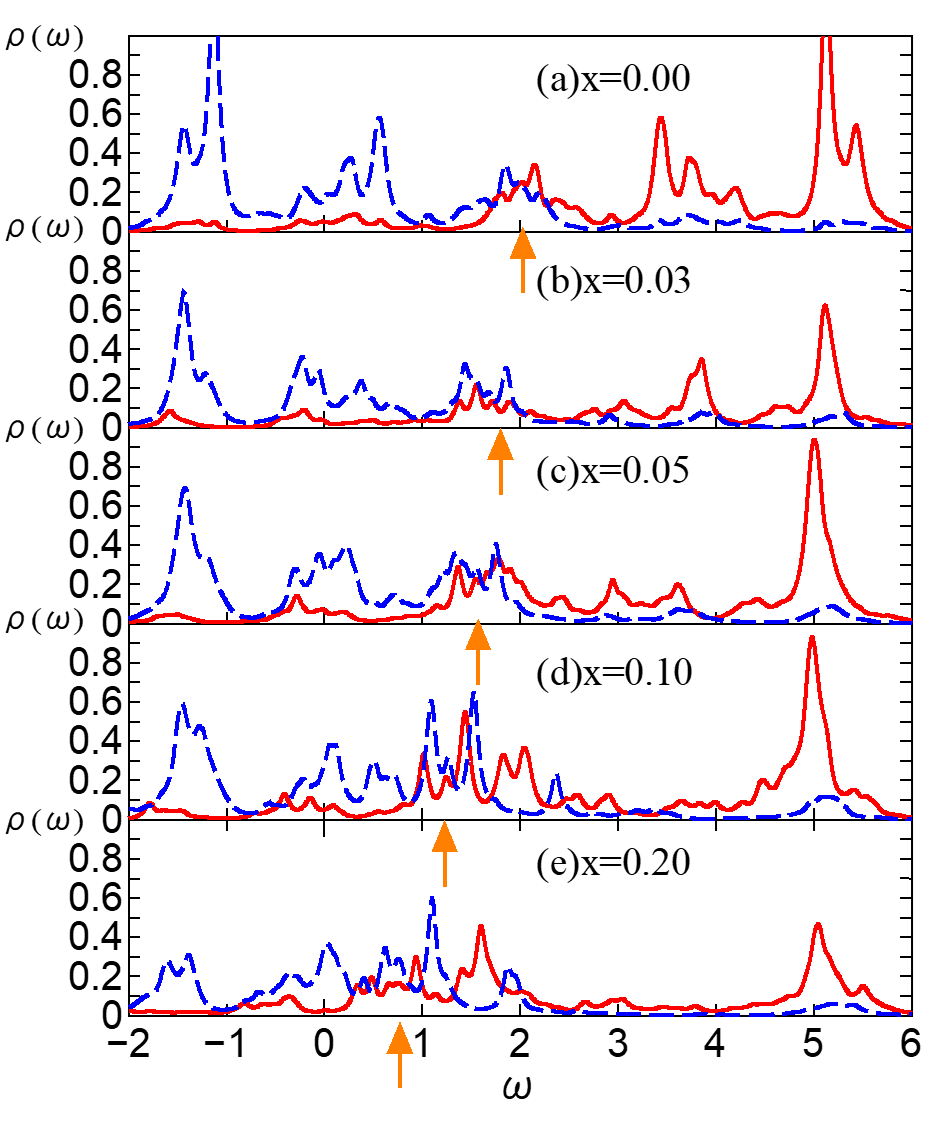}
\end{center}
\caption{\label{fig4} 
(Color online) The spectral function of $c$ electron $\rho_c(\omega)$ (solid lines) and that of $f$ electron  $\rho_f(\omega)$ (dashed lines) as functions of the frequency $\omega$ at $U=4$ for $x=0.00$ (a), $x=0.03$ (b), $x=0.05$ (c), $x=0.10$ (d) and $x=0.20$ (e), where the arrows indicate the chemical potential $\mu$. The results from $N_s =6$ and 8 are merged and plotted with Lorentzian broadening with a width of 0.1eV.}
\end{figure}

Figs.~4 (a)-(e) show the $\omega$ dependence of $\rho_{c}(\omega)$ and  $\rho_{f}(\omega)$ at $U=4$ for several values of $x$. When $x$ increases, the quasiparticle peak centered at $\mu$ shifts downward together with the increase in the band overlapping (see also Fig.~3 (b)). At the same time, the weight of the quasiparticle peak of $c$ electron and that of $f$ electron increase with increasing $x$ in proportion to $Z_c$ and $Z_f$, respectively, where the former weight is relatively smaller than the latter one as  $Z_c<Z_f$ as shown in Fig.~3 (a).

Finally, we discuss the effect of the hole doping $x$ on the band insulator for a large $U>U_c$. 
Figures~5 (a) and (b) show the $x$ dependence of $Z_c$ and $Z_f$ and that of $n_c$ and $n_f$ together with $E_{\rm gap}$ given in eq. (5) for $U=6$, where the $N_s$ dependence of the obtained results for $N_s=4$, 6 and 8 is quite small as seen in Figs.~1 and 3. Remarkably, the both $Z_c$ and $Z_f$ rapidly decrease with increasing $x$ from $Z_c=Z_f\sim 1$ for $x=0$ down to $Z_c=0.06$ and $Z_f=0.22$ for $x\sim 0.01$ above which the both $Z_c$ and $Z_f$ monotonically increase towards unity for $x=1$. As shown in Fig.~5 (b), $n_f$ monotonically decreases with increasing $x$ from $n_f\sim 1$ for $x=0$ down to zero for $x=1$ while $n_c$ increases from $n_c\sim 0$ for $x=0$ to $n_c\sim 0.13$ for $x\sim 0.3$ above which it gradually decreases down to zero for $x=1$. As the results f $n_c$ and $n_f$, the band gap $E_{\rm gap}$ monotonically decreases with increasing $x$ from $E_{\rm gap}\sim \Delta+U-W=2.5$ for $x=0$ to zero at $x\sim 0.16$ above which the band overlapping $-E_{\rm gap}$ monotonically increases up to $W-\Delta=3.5$ for $x=1$. We note that the system is not a band insulator for $0 \siml x \siml 0.16$ where $E_{\rm gap}>0$ but $\mu$ does not sit inbetween the gap in contrast to the case with $x=0$. 

\begin{figure}[t]
\begin{center}
\includegraphics[width=6.4cm]{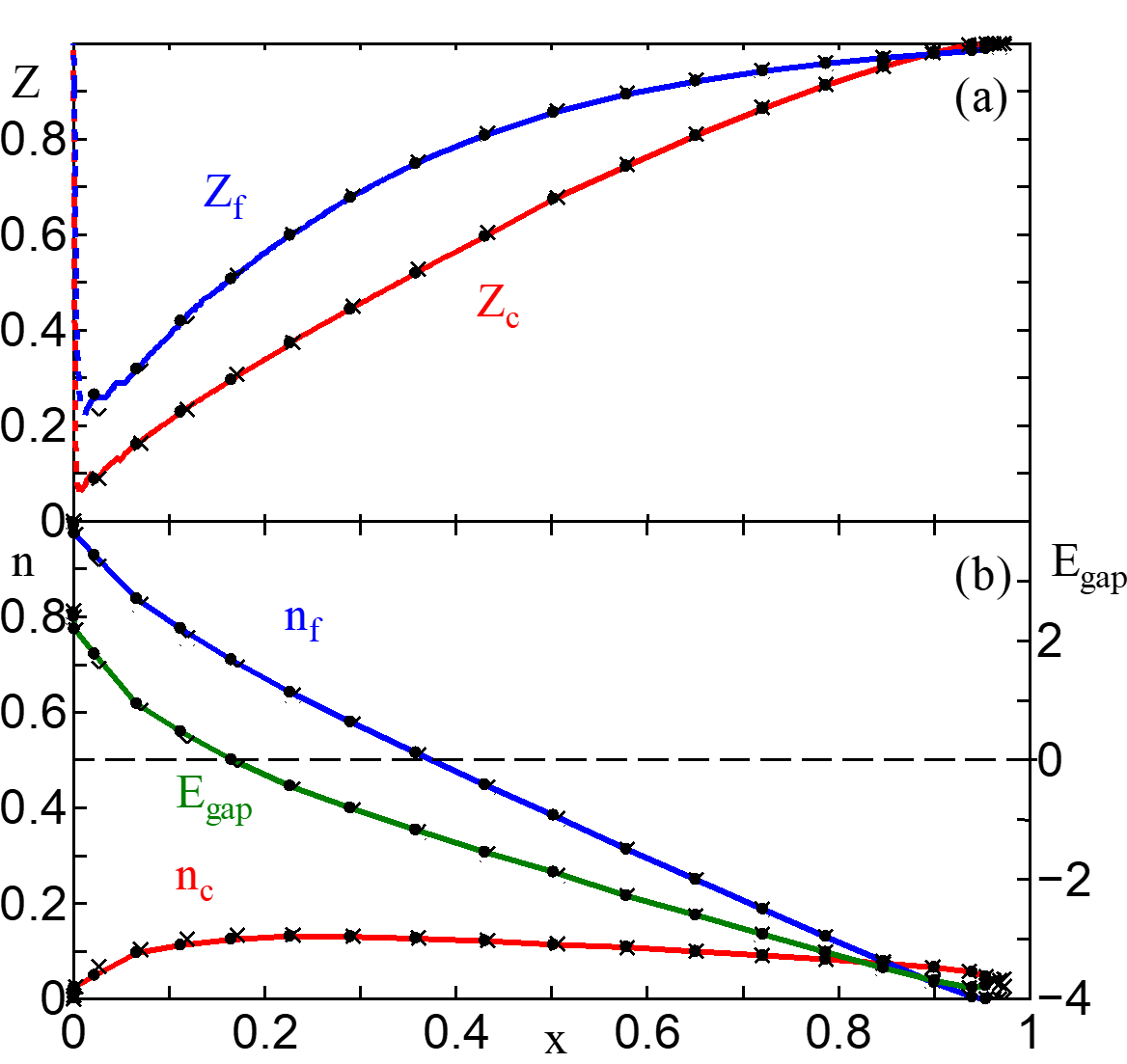}
\end{center}
\caption{\label{fig5} 
(Color online) The renormalization factor of $c$ ($f$) electron $Z_c$ ($Z_f$) (a) and the occupation number of $c$ ($f$) electron $n_c$ ($n_f$) together with the band gap $E_{\rm gap}$ as functions of the hole doping $x$ at $U=6$ for $N_s =4$ (solid (dashed) lines for metallic (insulating) state), $N_s =6$ (dots) and $N_s =8$ (crosses).}
\end{figure}

Figs.~6 (a)-(f) show the $\omega$ dependence of $\rho_{c}(\omega)$ and  $\rho_{f}(\omega)$ at $U=6$ for several values of $x$. In the case with a slight hole doping $x=0.01$, remarkable in-gap states consist of $c$ and $f$ electrons emerge above the valence band top as shown in Fig.~6 (b) and are responsible for a correlated semimetal with a large effective mass $(m^*/m)_{c(f)}=Z_{c(f)}^{-1}\sim 16.7$ (4.5) (see Fig.~5 (a)) as $\mu$ sits in the in-gap state. When $x$ increases for $x\simg 0.01$, the in-gap state shifts downward together with $\mu$ and the weight of the in-gap state of $c$ ($f$) electron increases together with the increase in $Z_c$ ($Z_f$) as shown in Figs.~6 (b)-(f) (see also Fig.~5 (a)). These features of the in-gap state in the doped band insulator for a large $U>U_c$ are similar to the in-gap states observed in doped Mott insulators\cite{Georges,Mutou}. On the other hand, for $0< x\siml 0.01$, the system is a thermally doped semiconductor with $\rho_{c(f)}(\mu)\sim 0$ and $\mu$ rapidly shifts downward with increasing $x$ so as to determine $x$ (see Figs.~6 (a) and (b)). Thus, a crossover from a thermally doped semiconductor to a correlated semimetal takes place at $x\sim 0.01$.

In summary, we have made a DMFT study on the spinless electron-hole two-band Hubbard model in the case with a small $c$-$f$ level splitting, where a transition from a semimetal ($U<U_c$) to a band insulator ($U>U_c$) is expected to occur at a large $U_c$ due to the Hartree shift. Actually we have found that the first-order Lifshitz transition takes place at $U_c \sim \mbox{bandwidth}$ from a correlated semimetal with $m^*/m=Z^{-1} \sim 5$ to an almost uncorrelated band insulator. When the hole is doped in the correlated semimetal with $U<U_c$, $Z$ has been found to monotonically increase with increasing the hole doping $x$ together with the increase in the band overlapping. On the other hand, when the hole is doped in the band insulator with a large $U>U_c$, a rapid crossover has been observed from a thermally doped semiconductor for $x\siml 0.01$ to a correlated semimetal with $m^*/m=Z^{-1} \sim 10$ for $x\simg 0.01$ where remarkable in-gap states consist of $c$ and $f$ electrons emerge above the valence band top and those weights increase with increasing $x$ together with the increase in $Z$ similar to the in-gap states observed in doped Mott insulators. Such in-gap states might be observed in metalized Ta$_2$NiSe$_5$ due to carrier doping\cite{Tsuchida} where the recent experiment of the Sommerfeld coefficient indicates a large effective mass\cite{Tsuchida2}. 

\begin{figure}[t]
\begin{center}
\includegraphics[width=6.4cm]{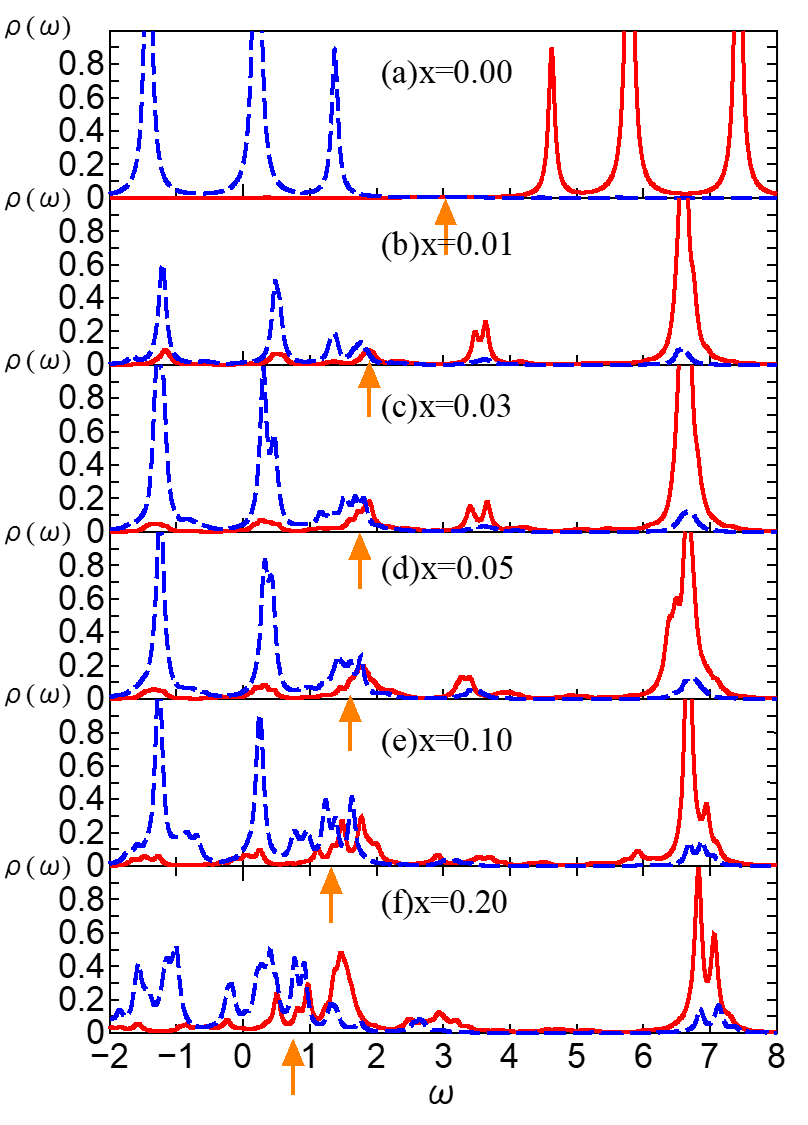}
\end{center}
\caption{\label{fig6} 
(Color online) Spectral function of $c$ electron $\rho_c(\omega)$ (solid lines) and that of $f$ electron  $\rho_f(\omega)$ (dashed lines) as functions of the frequency $\omega$ at $U=6$ for $x=0.00$ (a), $x=0.01$ (b), $x=0.03$ (c), $x=0.05$ (d), $x=0.10$ (e) and $x=0.20$ (f), where the arrows indicate the chemical potential $\mu$. The results from $N_s =6$ and 8 are merged and plotted with Lorentzian broadening with a width of 0.1eV.}
\end{figure}

\begin{acknowledgments}
The authors thank S. Tsuchida, Y. Hirose, R. Settai, T. Kaneko, K. Kuroki and T. Sekikawa for fruitful discussions. This work was partially supported by JSPS KAKENHI Grant Number 25K07201.  Numerical calculations were performed using the facilities belonging to the Research Center for Computational Science, Okazaki, Japan and the Center for Computational Sciences, University of Tsukuba. 
\end{acknowledgments}

\end{document}